\documentclass{elsart}
\usepackage{epsf}
\def\ds{\displaystyle}
\begin{document}
\begin{frontmatter}
\title{Experimental observations of a new information channal in nature
with aim of quartz resonators system}
\author{Yu.A.Baurov\thanksref{1}}
\author{V.A.Yakovenko, A.V.Komissarov,}
\author{V.G.Verzhikovski, A.Yu.Baurov}
\address{Central Research Institute of Machine Building, \\ 141070, Pionerskaya 4, Korolyov, Moscow region, Russia.}
\author{\fbox{A.A.Konradov}}
\address{Institute of Biochemical Physics of Russian Academy of Sciences (RAS), \\ 117977,
Kosygin Str. 4, Moscow, Russia.}
\author{T.A.Zenchenko}
\address{Institute of Theoretical and Experimental Biophysics of RAS, 
\\ 142290, Institutskaya 3, Puschino-on-Oka, Moscow region, Russia.}
\thanks[1]{baurov@mail.ru}
                                   
\begin{abstract}
The results of long-term (two-year) experimental observations of frequencies of
two quartz resonators, one of which is placed in special magnetic system that 
creates vector potential field, another (calibration one) is placed outside 
this system, have been presented. Changes with different periodicity: 24 h, 
high-definite 7 days complex-form period, 27 days and year periods were detected
 during the observation of differences of these quartz resonators frequencies. 
The tangents which have been drawn to a terrestrial parallel at the moment of 
near-daily observation of measured quantity minima form the basic, most powerful
 subset of directions of tangents having a sharp corner with dipole component of
 vector potential of the Sun' magnetic field in a range from $50^\circ$ up to $80^\circ$
 at annual rotation of laboratory around the Sun together with the Earth. 
Also there are three subsets of directions similarly drawn tangents fixed 
in physical space that coincide with similar subsets of tangents directions 
drawn in the same way to points of radioactive elements $\beta$-decay count 
rate minimum under the long $\beta$-decay research during daily and annual 
rotation of laboratory together with the Earth. It is necessary to point out 
that the form of the curves of quartzes frequencies difference changes in scales
 of the 7-day's period with reliability coincides next year, and observable 
minima of frequencies in co-phase points of space coincide to within several 
tens minutes. 
It is shown that the amplitude of the signal was $5\div10$ times lesser in the period from 20-th, July, till
20-th, January, than from 20-th, January, till 20-th, July. The results are in agreement with non-gauge theory of physical space formation in which potentials become observable and with a hypothesis about existence of anisotropic interaction of objects in  nature, caused by existence of fundamental vector constant - cosmological vector potential $\bf A_g$.
\end{abstract}

%{PACS 52.30, 12.60}
\begin{keyword}
quartz resonator, space anisotropy
\PACS 77.65.F, 12.60
\end{keyword}
\end{frontmatter}

\section{Introduction.}

In Refs.[1-6] the experiments are described in which, with the use of high-current 
magnets, torsion and piezoresonance balances, the new phenomena were for the 
first time detected. These phenomena may be interpreted as a new interaction of
 objects in nature, distinct from the known strong, electromagnetic, and 
gravitational interactions. The main distinctive property of this interaction 
is its spatial anisotropy. The new interaction is associated, from our viewpoint,
 with the existence of a new fundamental vectorial constant-cosmological vector
 potential $\bf{A}_g$ substantiated in Refs.[5,6] where the processes of formation 
of charge numbers of elementary particles are studied and hence, the gauge invariance 
is violated. Thus the potentials of physical fields acquire a physical sense in
 such a theory and become measurable on the set of formation of charge numbers 
of elementary particles (the electric charge, particularly). According to Refs.
 [4-6], the anisotropic interaction is nonlinear, nonlocal, and representable by a
 series in a change (difference) of some summary vector potential  $\bf{A}_{\Sigma}$
  at the location points of a "sensor" and "test" body. The summary vector 
potential includes the vectorial potentials from all the sources of currents 
and magnetic fields (for example, from the Sun $\bf{A}_{\odot}$, the Earth $\bf{A}_{E}$, 
the solenoids, etc). Therewith always $|\bf{A}_{\Sigma}| \le |\bf{A}_g| \approx 1,95\times10^{11} Gs\cdot cm$.

The nonlocality of the new interaction manifests itself in that its magnitude 
depends on the difference of values of $\bf{A}_{\Sigma}$ determined at the points
 of location of the "sensor" and the "test" body. If, for example, both are at 
the same $\bf{A}_{\Sigma}$, the new force will be zero [5,6]. The property 
described manifests itself as when the "sensor" and "test" body are "confluenced" 
at one point of space, so when the values of $\bf{A}_{\Sigma}$ are equal at 
different spatial points. The latter is easily seen from the analysis of the 
first term of expansion of the expression for the new force in terms of changes
in $\bf{A}_{\Sigma}$, i.e. $\Delta\bf{A}_{\Sigma}$.
\begin{equation}
F \sim {\Delta A}_{\Sigma}\frac{\partial{\Delta A}_{\Sigma}}{\partial x}
\end{equation}
Here $x$ -- is the spatial coordinate of the observable physical space $R_3$. 
The same anisotropic properties of the new interaction manifest themselves in 
a huge range of scales: from $10^{-17}cm$ up to $10^{28}cm$, as well as in various 
phenomena. For example, it is shown in Ref.[7] that the observed variations in 
the count rate of $\beta$-decay of radioactive elements $^{60}Co$, $^{137}Cs$ up
 to $\sim (0.7-0.8)\%$ when the tangent line to the earth's parallel of latitude
 drawn at the moment of observation of the minimum decay rate at the place of 
location of the laboratory during its rotation together with the Earth, passes 
through certain spatial directions. In Refs. [8,9] the results of long-term ($\sim 1$ year)
 experimental investigations of the heat release in plasma devices of plasma 
generator-type in dependence on the spatial orientation of their discharge 
currents, are presented. The experiments have shown that the maximum action of 
the new force is observed at each point of space along the generatrix of a 
cone formed by this generatrix around the vector $\bf{A}_g$ with an opening 
angle of $\sim (90-100)^\circ$ when the vector potential $\bf{A}$ of any current
 (magnetic) system is directed oppositely to the vector $\bf{A}_g$ at an angle 
of $\sim (130-135)^\circ$. At that the vector $\bf{A}_g$ has the following 
coordinates in the second equatorial system: right ascension $\alpha \approx 293^\circ\pm 10^\circ$,
 declination $\delta \approx 36^\circ\pm 10^\circ$.

The spatial directions found in the above mentioned work [7] and in the earlier
 works on investigating the variations of the rate  of $\beta$-decay of various
 radioactive elements [10-13], are aligned with the generatrix of the above 
mentioned cone (with an accuracy of $\pm 10^\circ$), or correspond to the 
direction of the vector $\bf{A}_g$. It should be noted that the variations 
in the $\alpha$ and $\beta$-decay rates of radioactive elements observed also 
in Ref.[14] may be qualitatively explained in the context of Refs.[5,6].

In the Refs. [5,6,15,16] while using quartz gravimeters Sodin and an attached 
magnet, the signals of unknown nature were detected, too. They cannot be 
explained within the scope of existing physical views. The axes of sensitivity 
of gravimeters (normal to the Earth) at the moments of appearance of signals, 
were also directed along the genetatrix of the above indicated cone with an 
accuracy of $\pm 10^\circ$. The amplitude of the signal reached up to 15 moon 
tides in some cases. The phenomenon discovered obtains an explanation on the 
basis of the new anisotropic force.

It should be noted that the attempts to detect a "nonelectromagnetic cosmophysical 
interaction" were undertaken not only in the listed works [5-7,15,16] but also 
earlier [17,18]. In Ref. [17], a system of quartz resonators was used for this 
purpose. In Ref. [18], when analyzing the results of long-term work of atomic 
frequency standards, an action of external disturbances was detected which had 
a gravitation-wave character (by the authors' meaning). In Ref. [18], in a run 
of measurements of differential frequency in the system of quartz resonators, 
there were revealed: a daily period; a correlation with the radioradiation of 
the Sun and with the gleam intencity of photobacterial culture; an advancing 
correlation with the geomagnetic activity. These results were also interpreted 
as having a gravitation-wave nature, but in the form of an abstract hypothesis 
without consideration of specific mechanisms.

The aim of the present work is an experimental study of a new quantum information
channel (QIC) in nature, associated with the presumed anisotropic interaction 
of objects with the physical vacuum, by a system of quartz resonators.

\section{The scheme of experiment.}

The experiment was carried out at TsNIIMASH during two years uninterruptedly. 
The first results (over six months) are presented in Ref. [19].

\begin{figure}[h]
\centerline{\epsfxsize=80mm\epsfysize=65mm\epsfbox{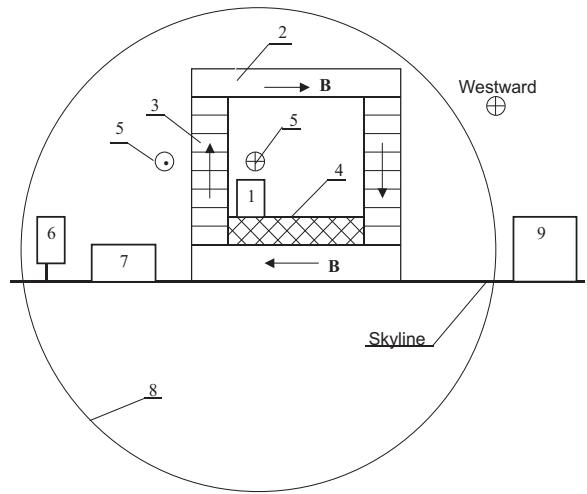}}
\caption{The experimental setup (all notes see in the text).}
\end{figure}

Figure 1 outlines the experimental setup. The element sensitive to the quantum 
information channel is the quartz resonator 1 of the quartz generator QG1 
placed within the magnetic system 2 of permanent magnets 3 creating a magnetic 
field of about 3500G in the magnetic circuit. The quartz resonator 1 is mounted 
on a conducting base 4 that allows it to be positioned in the space region with
 the minimal field of $B \sim 1$G. The magnetic system 2 is the source of the vector
 potential field 5 and of the gradient of the change in the modulus of a certain sum potential $\bf A_\Sigma$  (i.e., $\ds\frac{\partial\Delta A_\Sigma}{\partial x}$, see (1)) that comprises the above-mentioned cosmological vector potential Ag and the potentials of all natural and artificial magnetic sources.
System 2 acts as an amplifier of signals passing through the structure of physical vacuum. Such signals, i.e. any changes in $\Delta A_\Sigma$ , by themselves may have negligibly small amplitudes because of the immense cosmic scale of the regions of $\Delta A_\Sigma$  changes and, therefore, may be practically unobservable without the magnetic system 2.
The system registering the new signals includes a reference quartz resonator 6 of generator QG2 situated outside the magnetic system, about 1m away from QG1. Both quartz resonators and the electronic unit 7 are housed in a closed thick-wall metal chamber 8, which significantly screens the external electromagnetic interference and also acts as a thermal damper and shield against various extraneous influences (such as air convection). Beside the chamber is a personal computer (PC) 9 registering the experimental parameters.

\begin{figure}[h]
\centerline{\epsfysize=25mm\epsfxsize=85mm\epsfbox{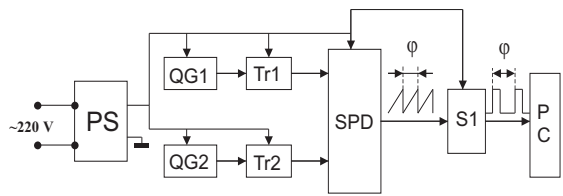}}
\caption{Block-diagram of the electronic devices.}
\end{figure}

Figure 2 is a block diagram of the electronic devices for monitoring the operation of the system. For precise measurement of the changes in the frequency $f_1$ of the QG1 quartz resonator, as indicated above, use is made of the reference generator QG2 with frequency $f_2$. The measured frequency difference $\Delta f = f_2 - f_1$ between QG2 and QG1 characterizes, in the framework of our concept, the change in  $A_\Sigma$  and hence the value of the new force. When the phase shift between the QG2 and QG1 oscillations reaches $\Delta\varphi = 2\pi$, the synchronous phase detector unit generates a pulse.
The sequence of pulses produced by the electronic scheme is fed to the PC where the time intervals between the pulse trains are determined to an accuracy of about $\sim 5\cdot 10^{-7}$s. To ensure such accuracy, the measuring board has a quartz resonator with a frequency of 10 MHz that sets a continuous sequence of time strobes of $10^{-7}$s duration.
The results of measurements and the current time are displayed at the PC terminal and simultaneously recorded on hard disk for storage and further processing.

\section{The results of experimental observations.}

%The experiment was carried out at TsNIIMASH %uninterruptedly throughout two years. The initial %results, over six months, have been published in %Ref. [93].

In Fig.3 the results of two-year experiment started on August 03, 2000, 
are shown. The heavy line corresponds to the behavior of $1/\Delta f$, 
the thin line does to the variation of temperature. The latter was measured daily in the morning by a mercury thermometer accurate to 0.1 degree which was positioned in the chamber with quartzs. It is seen from the Figure that both dependencies have very close yearly cycles.% denoted by a tick. 

\begin{figure}[h]
\centerline{\epsfysize=100mm\epsfxsize=120mm\epsfbox{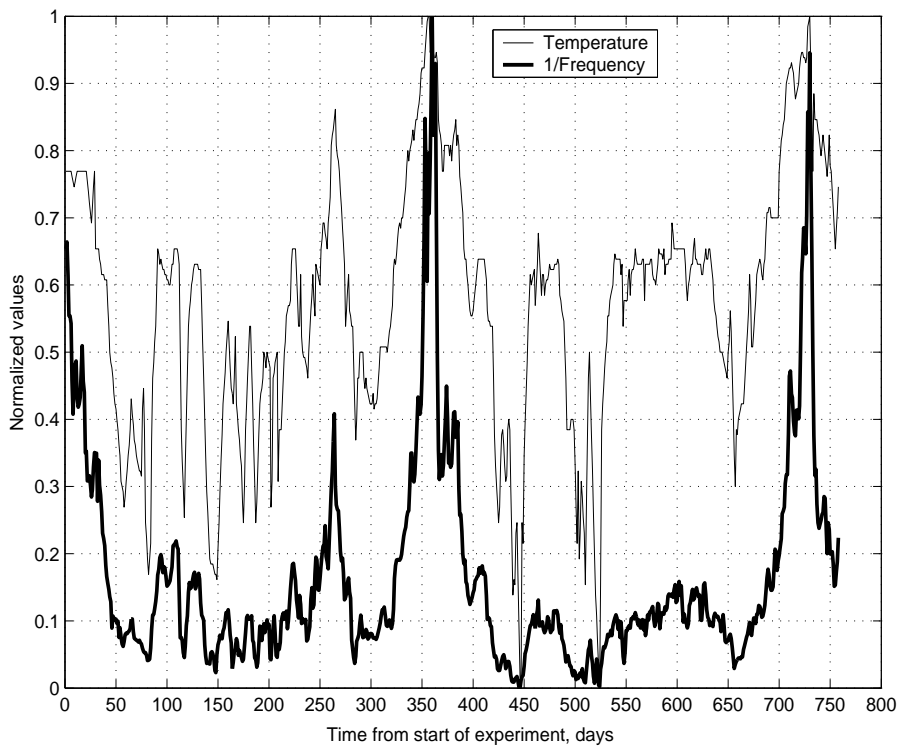}}
\caption{The results of two-year experiment started on August 03, 2000.
Bold line corresponds to the behavior of $1/\Delta f$, the thin line does 
to the variation of temperature.}
\end{figure}

The results of Fourier - analysis for the whole number of data in the hour's and daily ranges are presented in Figs.4-6. As evident from the Figures, a near-hour's, near -weekly, and near - 27-dayly periods stand out. 

\begin{figure}[ht]
\centerline{\epsfysize=90mm\epsfxsize=90mm\epsfbox{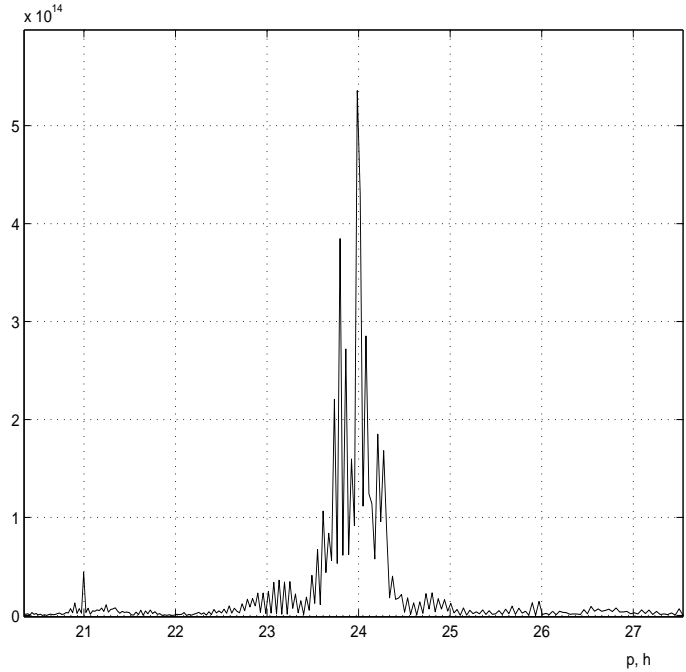}}
\caption{The results of Fourier-analysis for the whole number of $1/\Delta$f data 
in the hour's range.} 
\end{figure}
\begin{figure}[ht]
\centerline{\epsfysize=90mm\epsfxsize=90mm\epsfbox{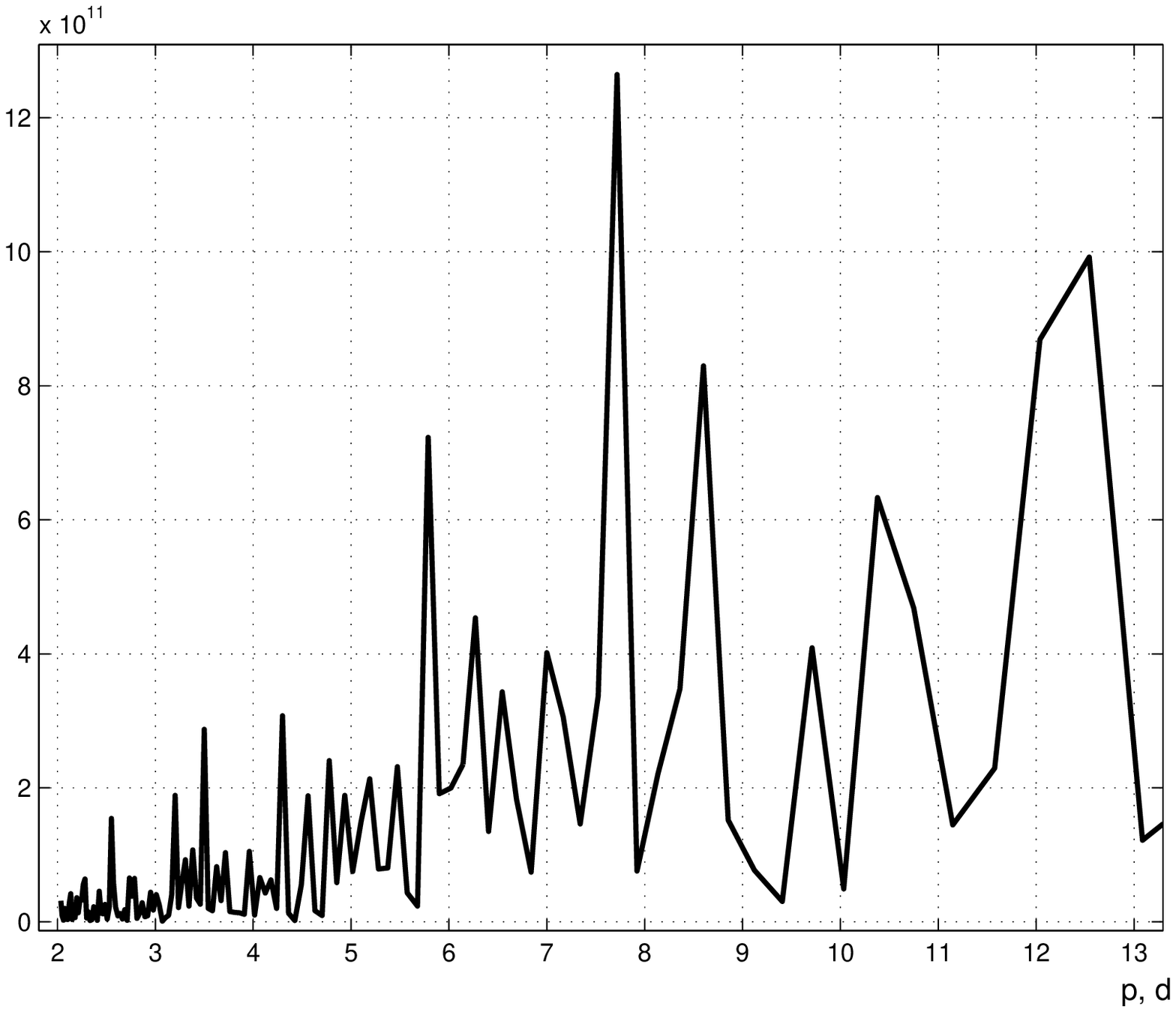}}
\caption{The results of Fourier-analysis for the whole number of $1/\Delta$f data
in the daily range.} 
\end{figure}
\begin{figure}[ht]
\centerline{\epsfysize=100mm\epsfxsize=100mm\epsfbox{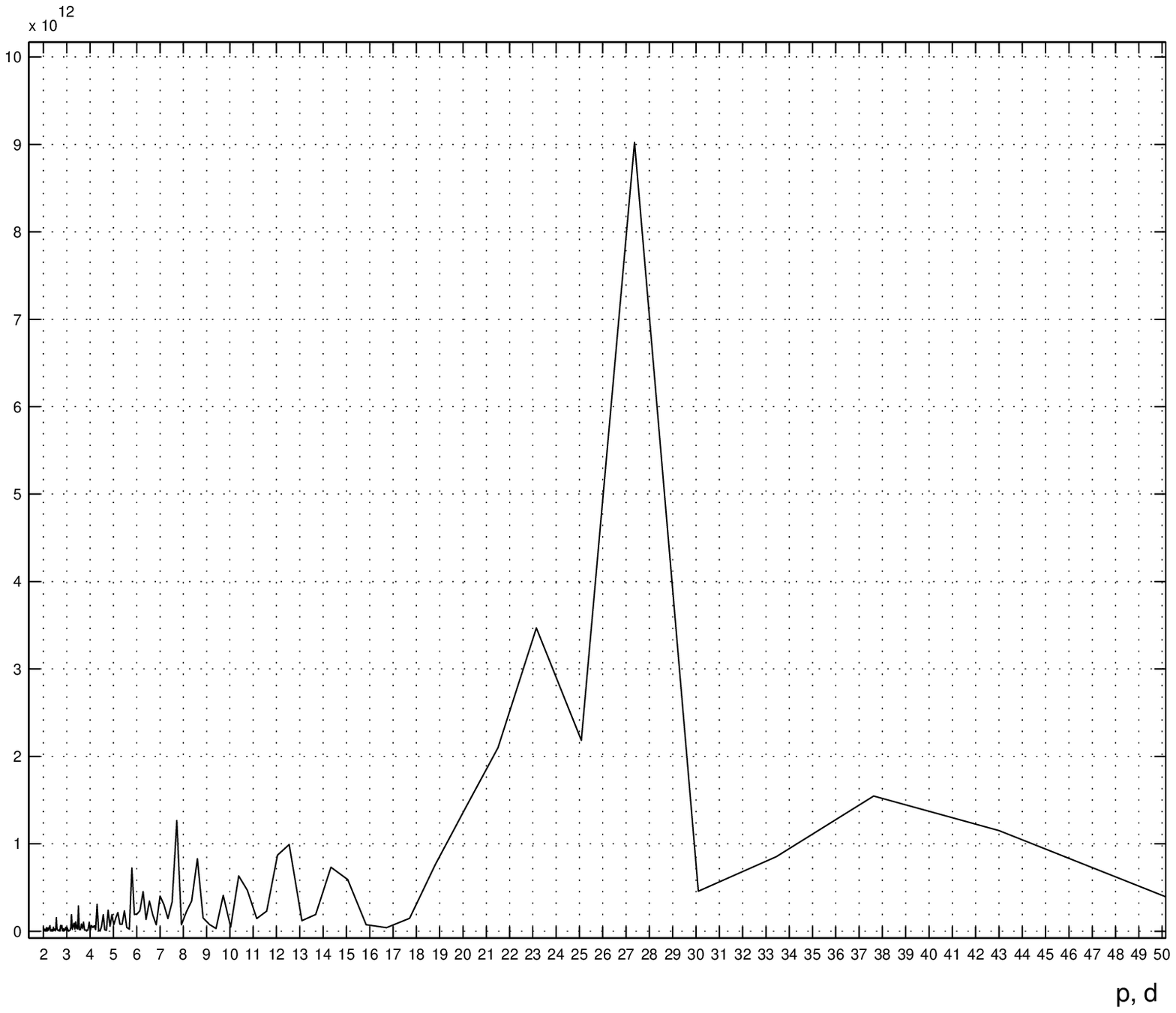}}
\caption{The results of Fourier-analysis for the whole number of $1/\Delta$f data
in the weekly range.}
\end{figure}

In Fig. 7, the changes in  $1/\Delta f$ (or, what is the same, changes in the 
phase run between two quartzs) for time intervals from 02.10.2000 till 10.10.2000
 and from 02.10.2000 till 09.10.2001, are shown. The striking similarity of Figures 
as to the form of signals so to the time points of minimum observations  
(with the maximum discrepancy between those being no more than several tens 
of minutes) argues for the non-randomness of the process in observation. 
The weekly complex-shaped cycle shown (Fig. 7) was observed in the course 
of the whole experiment. 

\begin{figure}[h]
\centerline{\epsfysize=200mm\epsfxsize=140mm\epsfbox{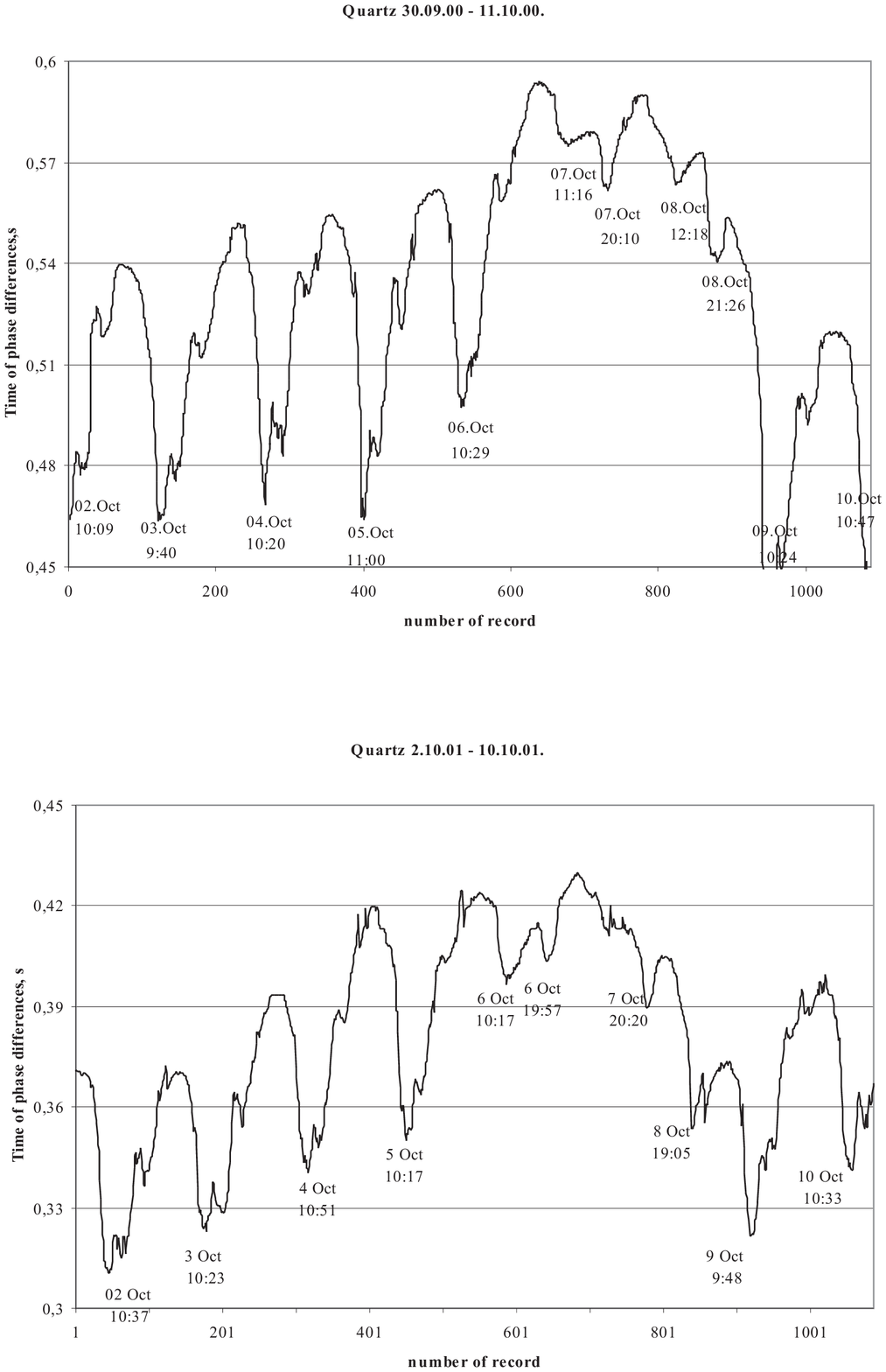}}
\caption{ The changes in the phase run between two quartzs (1/$\Delta f$) for time intervals 
from 02.10.2000 till 10.10.2000 and from 02.10.2000 till 09.10.2001.}
\end{figure}

At first sight it would seem from the Fig.3 that the phenomenon observed was 
caused by temperature oscillations. To check that possibility, we have taken 
special measurements of temperature each hour during two and half days. 
The results are given in Fig.8. It is clear that the near-daily oscillations 
of  $1/\Delta f$ cannot be explained by the changes in temperature. 

\begin{figure}[h]
\centerline{\epsfysize=80mm\epsfxsize=90mm\epsfbox{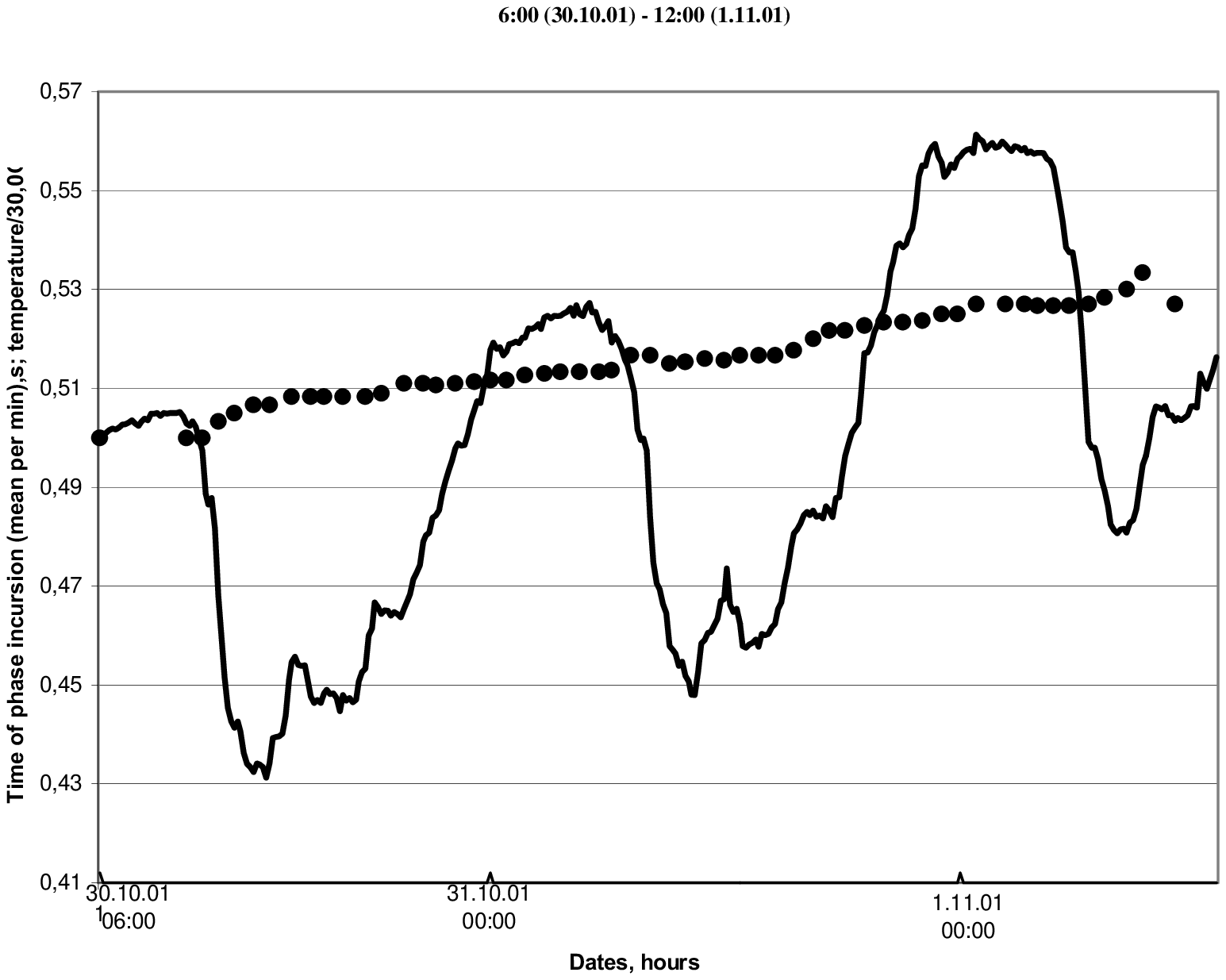}}
\caption{The results of measurements of temperature each hour during two and half days
(6.00 30.10.01. - 2.00 01.11.01.) and the changes in $1/\Delta$f at the same time.}
\end{figure}

As is seen from Fig.8, the response of the system of quartz resonators to the 
temperature variation is not limited by a linear dependence. More or less 
regular deviations from the regression line are observed, which is evident 
also from Fig.3. To study the response of the system, it is shown in Fig.9 
that the linear part of the temperature (t) dependence of  $\Delta f$ can 
be approximated by the regression f = $5,49 - 0,25*t$. The results subtraction 
of this dependence from the values of two-year observation of changes in  
$\Delta f$, are presented in Fig.10. The more than two - fold outliers of  
$\Delta f$ above the standard deviation, are numbered.

\begin{figure}[h]
\centerline{\epsfysize=100mm\epsfxsize=100mm\epsfbox{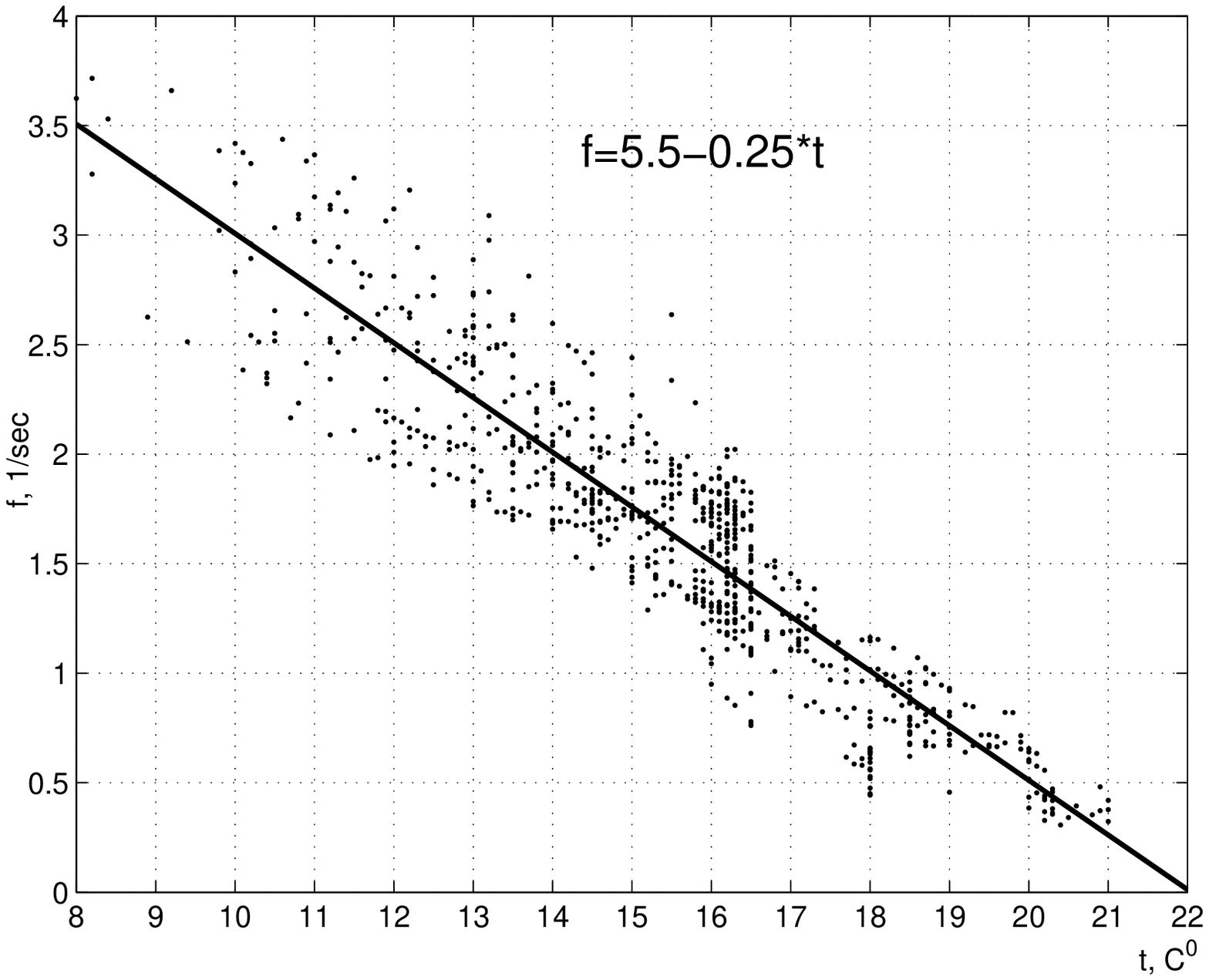}}
\caption{Linear approximation of $\Delta$f(t) dependences: f = 5,49 - 0,25*t.}
\end{figure}

In Fig.11 the trajectory curve of Earth's motion around the Sun is shown. 
At a stable position of the Sun's dipole this curve is very close to the line 
of the vectorial potential of the Sun's magnetic dipole. Shown are also the 
trajectories of the motion of the laboratory during its rotation together 
with the Earth (those are very close to the lines of the vectorial potential of
 terrestrial magnetic field) as well as the positions of the laboratory at 
time points of observation of minimum  $1/\Delta f$. Large crosses denotes 
in Fig.11 the spatial location of the outliers numbered in Fig.10. As is seen 
from Fig.11, the time interval between the outliers (2,3) and (6,7) practically
 repeats after the year. The rest of outliers (1,4,5,8) will be discussed below.
 In the present paper we consider in more detail only several time intervals as to observation of changes in  $1/\Delta f$.

 \begin{figure}[h]
\centerline{\epsfysize=100mm\epsfxsize=120mm\epsfbox{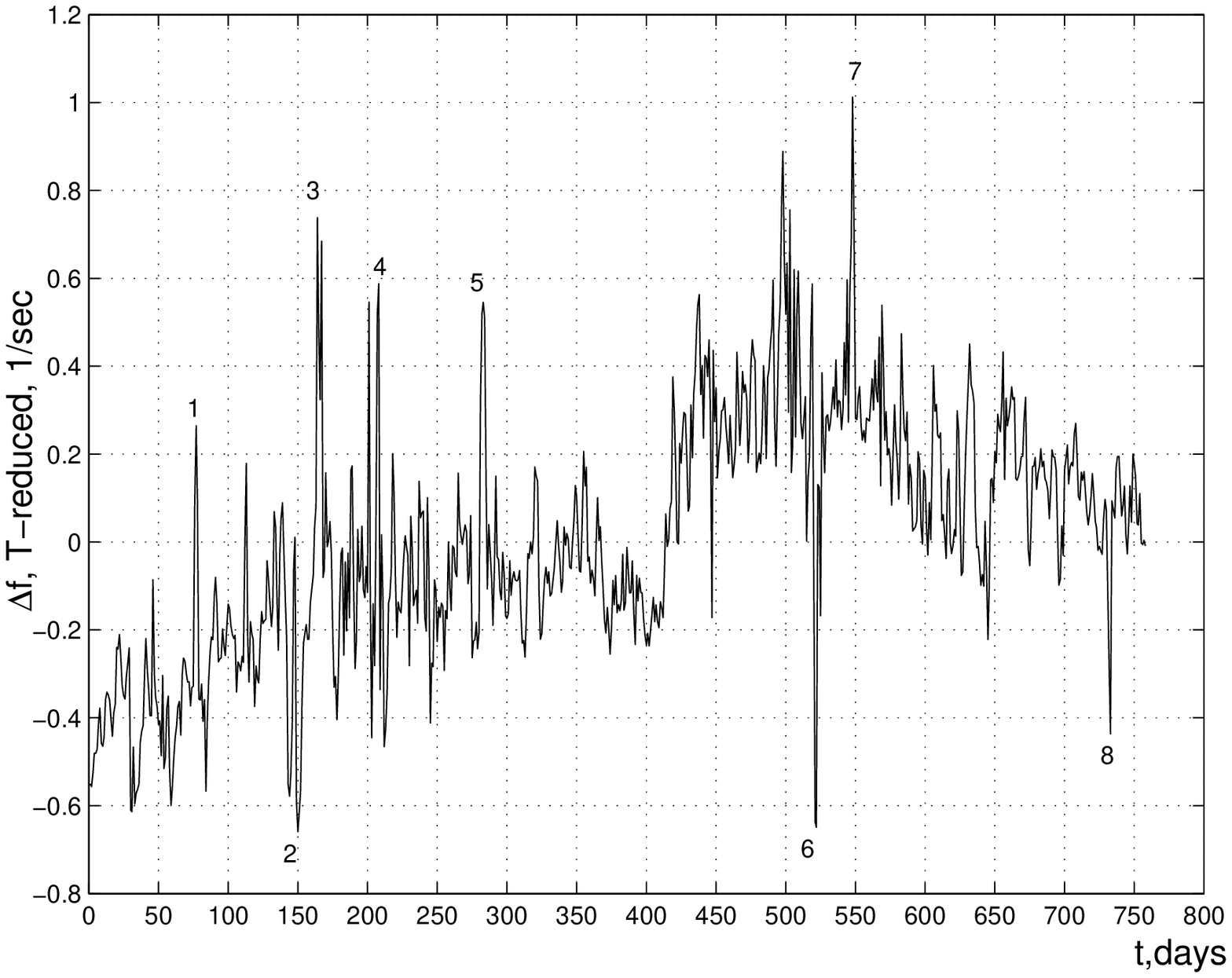}}
\caption{Changes in $\Delta$f reducing by linear temperature regression f = 5,49 - 0,25*t (Fig.9) during two-years experiment. More than two-fold outliers of $\Delta f$ above the standard deviation, are numbered.}
\end{figure}

\begin{figure}[h]
\centerline{\epsfysize=95mm\epsfxsize=100mm\epsfbox{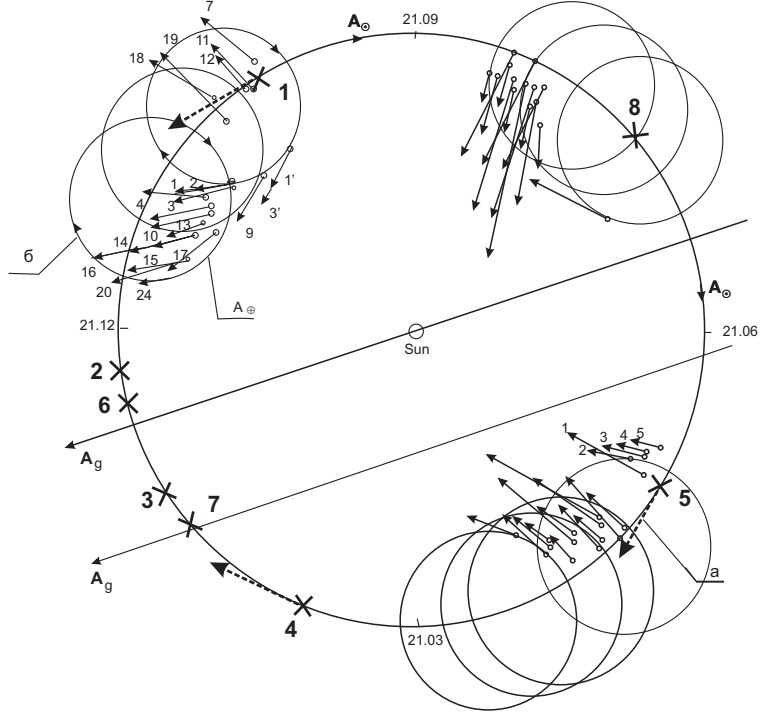}}
\caption{The directions of tangent lines to the Earth's parallels of latitude (practically, lines of vector potential of Earth's magnetic field) at places where the minimum values of $1/\Delta f$ were observed (practically, directions of a new force action) as well as the positions of the laboratory at 
time points of this observations. Large crosses denotes in Fig.11 the spatial location of the outliers numbered in Fig.10.
$\bf A_\oplus$ - direction of vector potential of Earth's magnetic field;
$\bf A_\odot$, - direction of vector potential of Sun magnetic field;
$\bf A_g$ - direction of cosmological vector potential;
$\circ\!\rightarrow$ -- placements of quartz resonators, corresponding to the time when the minimum values of $1/\Delta f$ were observed (and the directions of a new force action at this moments);
b - trajectories of the motion of the laboratory (quartz resonators) during its rotation together 
with the Earth; a - the trajectory curve of Earth's motion around the Sun.}
\end{figure}

\begin{figure}[h]
\centerline{\epsfysize=100mm\epsfxsize=140mm\epsfbox{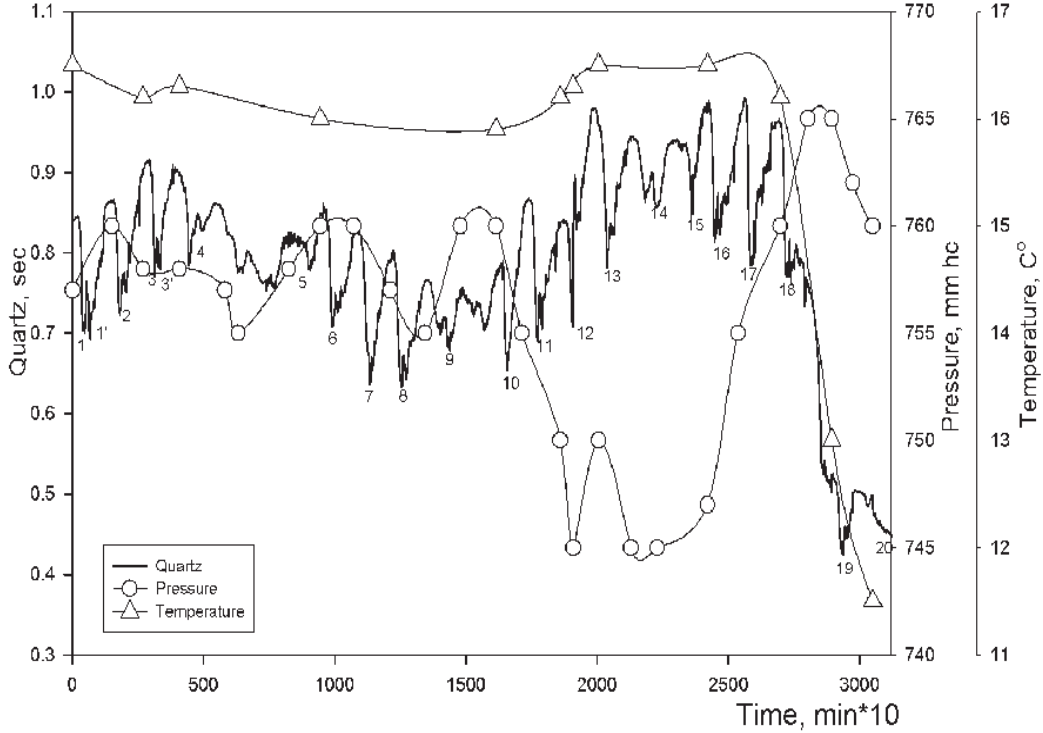}}
\caption{Changes in $1/\Delta$f, temperature and pressure from 01.11.2000. to 26.11.2000.}
\end{figure}

\begin{figure}[h]
\centerline{\epsfysize=100mm\epsfxsize=100mm\epsfbox{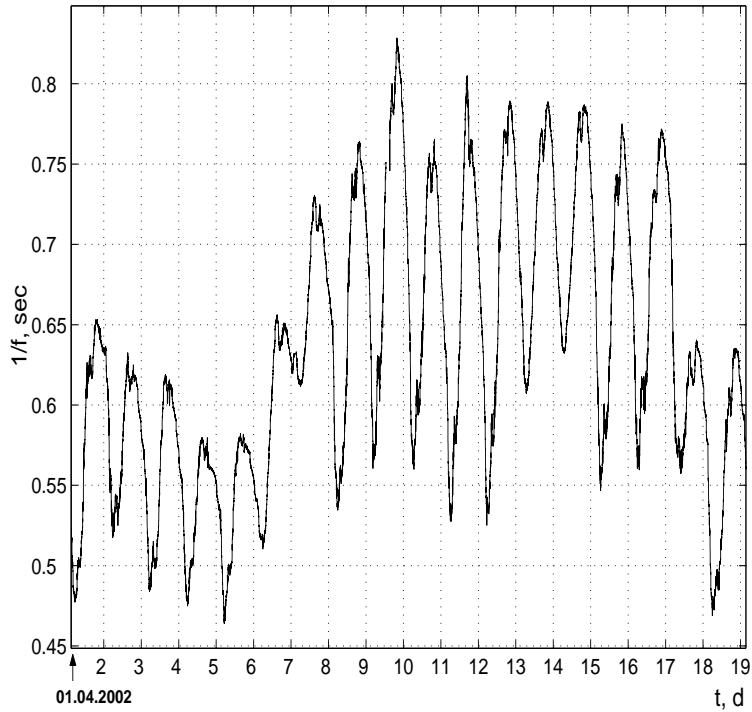}}
\caption{Changes in $1/\Delta$f from 01.04.2002. to 19.04.2002г.}
\end{figure}
\begin{figure}[h]
\centerline{\epsfysize=100mm\epsfxsize=100mm\epsfbox{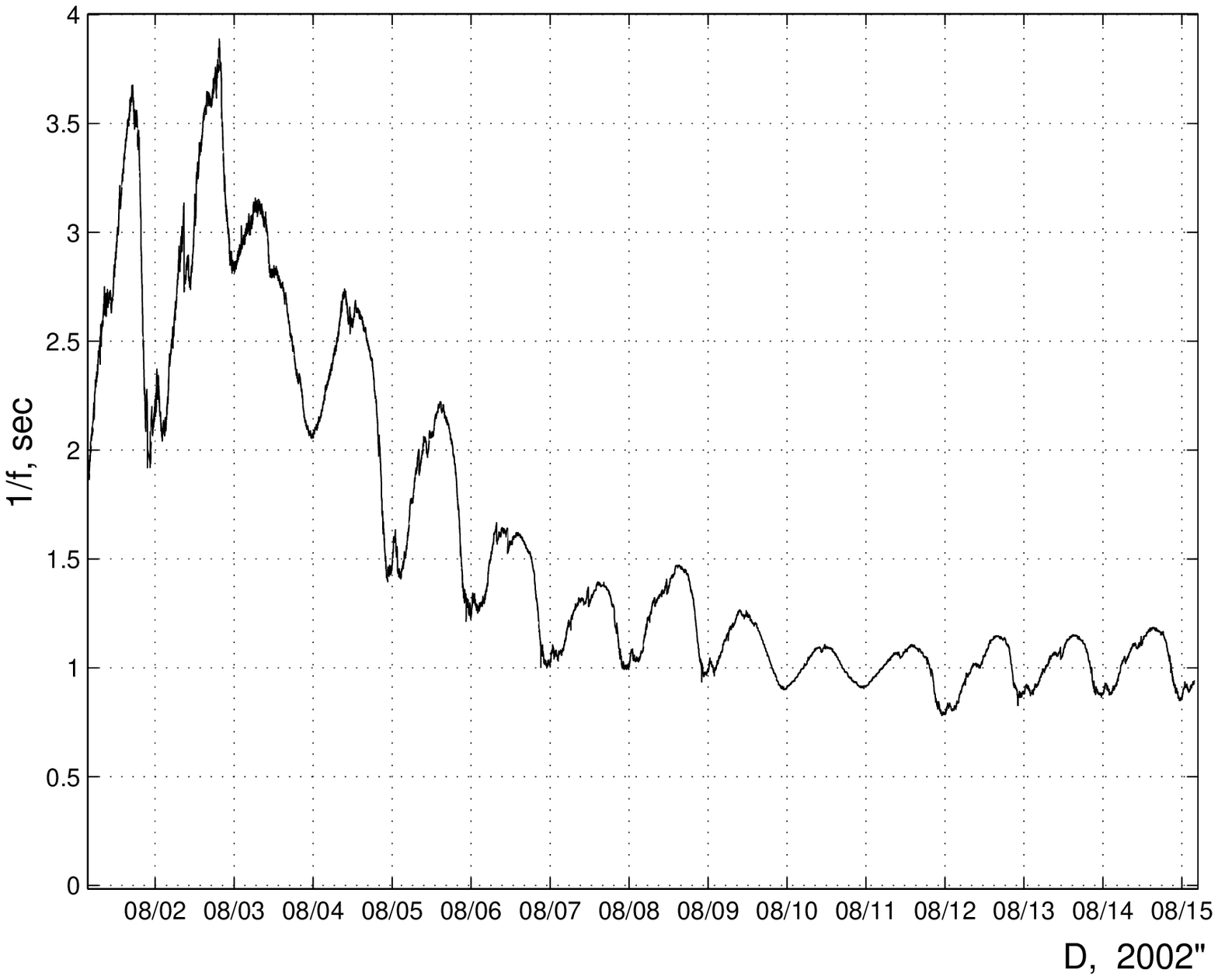}}
\caption{Changes in $1/\Delta$f from 01.08.2002. to 14.08.2002г.}
\end{figure}

 In Fig.12 are shown such changes (more exactly, the phase run in seconds) within the span from 01.11.2000 till 26.11.2000, as well as the changes in temperature and pressure over the same period. As is seen from the Figure, the change of pressure does in no way influence upon the result in consideration, i.e. upon the change in  $1/\Delta f$. The minimums observed are numbered. The directions of tangent lines to the Earth's parallels of latitude at places where the minimum values of  $1/\Delta f$ were observed, are shown in Fig.11. The said parallels are practically coincident with the lines of vectorial potential of the Earth. As is seen from the Figure, the tangent lines form three well defined spatial directions with an accuracy of $\pm 10^\circ$. In Fig.13 are shown the changes of $1/\Delta f$ over the period from 01.04.2002 till 19.04.2002. The similar tangents can be seen in Fig.11. As is evident from this Figure, the directions of arrows are near - alike (with the spread of $\pm10^\circ$). The changes of $1/\Delta f$ from 01.08.2002 till 14.08.2002 are given in Fig.14, and the corresponding tangents are in Fig.11. Most of the arrows have the same but new direction with the spread of $\pm10^\circ$ . One arrow is directed similarly to those (7,11,12,18,19) from the above shown experimental series (from 01.11.2000 till 26.11.2000). To obtain the general estimation of the role of the technogenic factor (in the workdays), in Fig.15 are presented the changes of  $1/\Delta f$ observed in the holidays from 30.04.2002 till 06.05.2002. The tangent lines drawn in the above manner at the instants of time when the correspondent numbered minimums of $1/\Delta f$ were observed, are presented in Fig.11. It is seen from that Figure that the tangents have approximately the same common direction. That has an angle of $\sim 30^\circ$ with the common direction in the series of measurement of  $1/\Delta f$ changes over the period from 01.04.2002 till 18.04.2002. An analysis of directions of above discussed tangents have shown that the major set of them is always directed at an angle of $50^\circ\div80^\circ$  to the line of the vectorial potential of the Sun $A_\odot$. That is, this set of directions is constantly turned in accordance with the motion of the Earth along its trajectory around the Sun. 
The less representative set of directions of tangent lines considered simply does not respond to the motion of the Earth and is always positioned on the above mentioned cone of direction of the new force. In the process of motion around the Sun, the basic set of tangents practically coincides with the directions of the axis or the generatrix of the cone considered during a third of the year (see Fig.11). It is notable that the vector ${\bf A_g}$ is directed along the axis of the cone in consideration.
At the period from 11.10.2002 till 26.10.2002, the influence of the magnetic system on the change of $1/\Delta f$ was investigated. The magnetic system (2) in Fig.1 was withdrawn from the chamber but the quartzs remained in the same position. The general character of the signal was shown to be not changed, it corresponded to that in Figs.12-15, but the amplitude of the signal was 
25\% lesser. The influence of the distance between quartzs on the character of signal in the absence of magnetic system, was investigated, too. The quartz resonator 6, remaining in the same plane, was connected to the quartz resonator 1 (Fig.1), and the character of changes in  $1/\Delta$f was observed over the period from 26.10.2002 till 06.11.2002. In the result, the amplitude of the signal was decreased by further 25\%, i.e. the change in  $1/\Delta$f was halved as compared with the basic experiment. In Fig.16 the sequence of observation of the time instants corresponding to minimum $1/\Delta$f in dependence on the starting point of the experiment, is presented. As is seen, the time points of minimum observations intersect the regression line. The distinction between the obtained distribution and the regression line was calculated to be confident at the significance level at least $P<0.001$ using run test [21]. This phenomenon was repeatedly observed in the experiment.

\begin{figure}[h]
\centerline{\epsfysize=95mm\epsfxsize=100mm\epsfbox{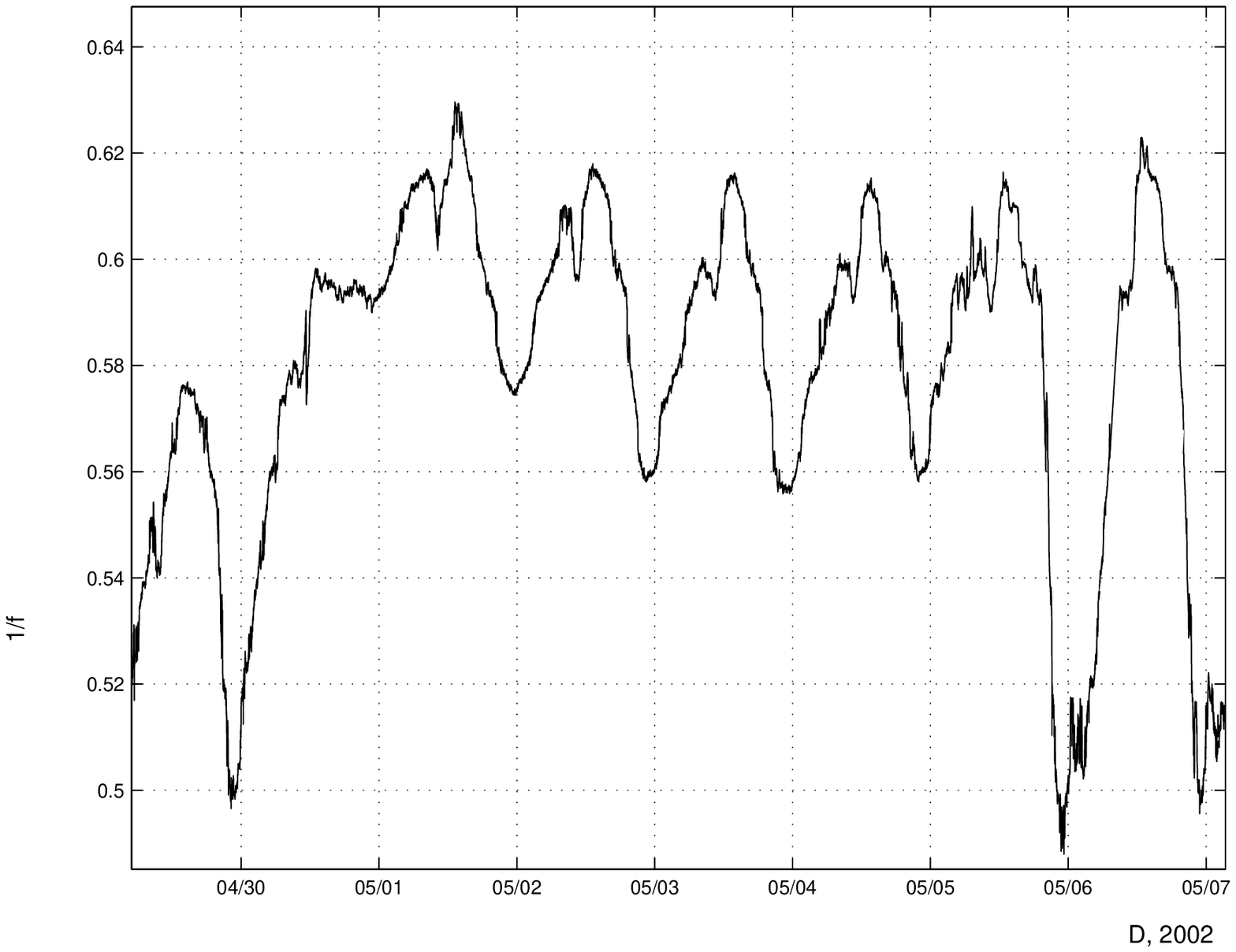}}
\caption{Changes of $1/\Delta f$ during the weekend and holidays (from 30.04.2002. to 06.05.2002).}
\end{figure}
\begin{figure}[h]
\centerline{\epsfysize=100mm\epsfxsize=100mm\epsfbox{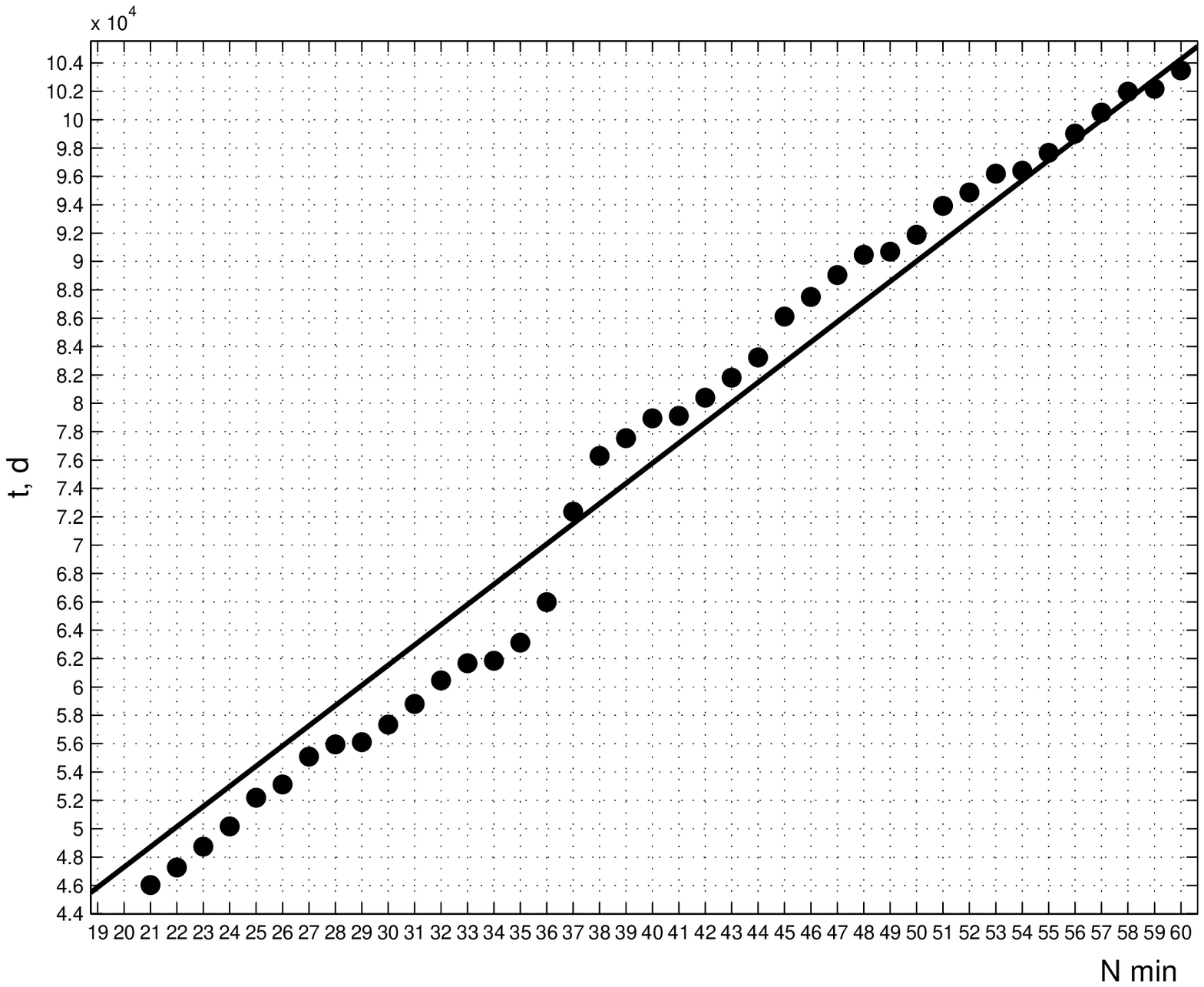}}
\caption{The sequence of observation of the time instants corresponding to minimum $1/\Delta$f in dependence on the starting point of the experiment and the regression line.}
\end{figure}

\section{The analysis of experimental errors.}

The errors of determining the direction of the vector $A_g$ and the properties of QIC with the aid of a system of quartz resonators can be subdivided into systematic and random (statistical) ones. The systematic errors are those due to the influence of changes in temperature, atmospheric pressure, supply-line voltage, magnetic field, and possible of the human factor. The statistic errors are caused by noises in the quartz resonators and in the electronics, by exactness of instrumentation (computer, voltage regulator, etc.), inaccuracy of geometric constructions, incontrollable electromagnetic noise, and power induction.
Consider the systematic errors. As is seen from Fig.3, the change in temperature plays an important role in the change of signal shapes during many-days observation of that. Nevertheless, due to the nonlinear response of the system of quartz resonators to the change of temperature (see Fig.3), we succeeded, with the use of a special mathematic procedure (Fig.9), in identifying its influence on the valid signal and obtaining the result shown in Fig.10. The analysis of this result is given below. To show the independence of the shape of the near-daily signal from the temperature, we measured its changes hourly during the course of two and half days with the aid of an exact mercury thermometer (having an accuracy of  $\pm 0,1^\circ С$ at the place of the system of quartz resonators). The results of that experiment are given in Fig.8. As is seen from this Figure, the shape of temperature does not correlate with that of the basic signal and therefore cannot be used for explanation of the main result obtained - i.e. the shape of change of phase difference of two quartz generators in the course of a day as well as the arrangement of the set of directions of tangents to the Earth's parallels of latitude drawn at the time instants of minimum phase differences of quartz generators, in the range of angles between 50 and 80 degrees relative to the vector $A_\odot$ (Fig.11). The atmospheric pressure was shown to have practically no influence on the experimental results (see Fig.12). Consider a systematic error connected with possible voltage deviations in the system of powering quartz generators. This error was investigated in the range of voltages between 180 and 230V at the nominal supply-line voltage of 220V. It was shown that the voltage regulator in the power unit of electronic equipment was not beyond the tolerance limits of stabilization. The voltage of the power unit was equal to 5V$\pm$0.025V which did not led to the change in the phase run time more that 0.5\%. The magnitude of magnetic field below of 2000Gs is shown to act, even if insignificantly ($10^{-6}$f), on the change of frequency of quartz resonators [20]. There fore the magnetic system used in the experiment considered was constructed in such a manner that the field magnitude at the location of quartzs to be no more that 1Gs and hence cannot act upon the deviation of $1/\Delta f$ more than 1\%.
The periodicity of the main signal ($\sim$24 hours), appearance of the major part of minimums at morning hours (near 10 hs), and the presence of a seven-day period seem to indicate to the possible role of the human factor in the experimental results. However those obtained in holydays (not workdays!) do not confirm the role of that factor (see Fig.15). As is seen from the Figure, the periodicity and the shape of  $1/\Delta f$ - changes are the same in holydays and in workdays.
Consider the random errors now. As was above said, the frequency of thermostable quartz generators remained constant with an accuracy of $10^{-7}$ at the standard frequency of quartzs $10^{6}$Hz. Moreover, because of differential connection of those the accuracy of measurements of  $1/\Delta f$ was even more than $10^{-7}$. Thus the noise in QG could not stand out above some percents of the amplitude of  $1/\Delta f$-deviation over the near-daily period.
The use of PCs for processing and recording the experimental results also leads to systematic (drift of clocks of PC's quartzs) and random errors (malfunctions of PCs and of power units). The drift of clocks (as of thermal so of other origin) leads to an insignificant error in time referencing of results of measurements of quartzs' frequency difference. That drift was equal, on the average, to no more than several minutes per month and was corrected by PC-operator according to the signals of Moscow time. PC was powered from the same stabilized power-line as the QG-system.
It is very difficult to explain the results of the experiment by the electromagnetic noise and corresponding stray pick-up since the obtained periodicity (24-hour's, weekly, 27-daily, and yearly periods) completely contradicts to such an assumption. Nevertheless it should be noted that in the vicinity of  $1/\Delta f$-minimums there were frequently observed outliers with duration between $\sim0.3\div1$ sec and with amplitudes of tens times more than the periodic  f-oscillations of not yet known nature. The geometric constructions in Fig.11 were made with the accuracy of $\approx 1^\circ$  and could not influence on the final conclusions as to the results of the whole experiment.

\section{Discussion of experimental results and some conclusions.}

Consider the periods of  $1/\Delta f$ observed in the experiment. As was said above, we failed in attempts to explain the near-24 hour's period in variations of $1/\Delta f$ by the thermal factor. The fluctuations of supply-line voltage (see above) also cannot be used for explanation of the said period. As the tangents to the Earth's parallels of latitude drawn for time instants of minimum $1/\Delta f$ observations always form nearly the same angle of $50^\circ\div80^\circ$  with the direction of the vectorial potential of the dipole component of the Sun's magnetic field, we may affirm that, in all likelihood, the influence of the process of addition of vectorial potentials of the Sun's and Earth's magnetic fields is observed here because at those extremum points the angle between the said potentials is equal to $100^\circ\div130^\circ$  in the course of the total two-yearly experiment and hence the changes in fundamental scales as well as, accordingly, in dimensions of crystalline lattices associated with the changes of $A_\Sigma$, are always much the same at said points. The near-weekly period observed is connected, in all probability, with some unknown technogenic factor since that cannot be explained by the variations of voltage (which is sufficiently stabilized), but it is seen at the same time from Fig.7 and Fig.15 that the "block" in the near-weekly period (its last three days) starts always or form Friday or from the day before holyday. The 27-daily period can be explained, as in Ref.[12], by the period of rotation of the Sun and of vectorial potential of Sun's magnetic field since in this case changes in ${\bf A_\Sigma}$ , fundamental scales, and crystalline lattice dimensions take place. The yearly period is also explained by the "influence" of the vector $A_\odot$ on the vector $A_\Sigma$ during the yearly rotation of the Earth around the Sun.

In all likelihood, the outliers 1,4,5 shown in Fig.10 are not random since the corresponding tangents to the line of vector $A_\odot$ drawn at time points of observation of those outliers, are aligned with the axis of the cone of action of the new force (outlier 1) or with the generatrix of this cone (outliers 4,5) (see Fig.11 and [9]). The outlier 8 (Fig.10) is practically opposite to the outliers 2,3 and 6,7 which were repeatedly observed after the year and therefore can be also explained by the "influence" of  $A_\odot$ on $A_\Sigma$, i.e. by not yet known law of addition of said vectors.

Before the experiments the authors anticipated that the magnetic system will act as a peculiar amplifier of signals of new nature as was observed earlier in the experiments with the gravimeter [15,16]. However the influence of the magnetic system was found to be less significant (%амплитуда сигнала изменилась на 
no more than 25\% changes of signal amplitude).
Therefore the system of quartzs in consideration may be thought to respond, most likely, not to the gradient of $\Delta A_\Sigma$  but simply to the change of $\Delta A_\Sigma$  in time during the rotation and motion of the Earth being accompanied by the smooth variation of fundamental scales at an insignificant value of the gradient $\frac{\partial\Delta A_\Sigma}{\partial x}$.

According to Refs.[5,6], $\Delta A_\Sigma$ = 0 at full junction of quartzs in one region of space, and hence the amplitude of the signal should not be changed. At the maximum convergence of quartzs in the experiment, the amplitude of the signal was halved in magnitude, which may be reasonably explained by approximation of  $\Delta A_\Sigma$  to zero. Really two quartzs can by no means be brought to a point.

Thus the analysis of the results of the experiment carried out shows, that the signals of new nature were detected. These results, for the most part, cannot be explained on a basis of commonly acknowledged standard physical views but may be qualitatively interpreted with the aid of the byuon model of formation of the surrounding physical space [5,6].

%Thus the analysis of the experiment carried out shows that the signals of new nature are detected. The bulk %of the results cannot be explained on the basis of recognized physical views but may be qualitatively %rationalized on the basis of the byuon model of formation of the surrounding physical space [5,6].

\section{Acknowledgments.}

This paper is written in memory of our colleague, the mathematician A.A.Konradov, passed away prematurely at the peak of his greative ability.

In conclusion, the authors would like to thank the known astrophysicist, doctor
 of physico-mathematical sciences, professor B.M.Vladimirsky (Krym Observatory,
 Ukraine) as well as all participants of the seminar "Cosmos and Biosphere" 
(Krym, Partenit, 2003) for the fruitful discussion of the results presented 
in this paper.

%Литература

%%%%%%%%%%%%%%%%%%%%%%%%%%%%%%%%%%%%%%%%%%%%%%%%%%%%%%%%%%%%%%%%%%%%%

\end{document}